\documentclass[preprint,aps]{revtex4}
\usepackage{amsmath,amsfonts,amssymb,bm,txfonts}
\usepackage{mathtools}
\usepackage{braket}
\usepackage[dvipdfmx]{graphicx}
\usepackage[dvipdfmx]{color}
\usepackage{here}
\newcommand{\half}{\frac{1}{2}}

\begin{document}

\title{Nonequilibrium Steady States in the Weakly-Coupled XXZ Model}
\author{Shimpei Senda, Tatsuro Sunami, Yuto Matsumoto, and Ayumu Sugita}
\email{a.sugita@omu.ac.jp}
\affiliation{Department of Physics and Electronics, Osaka Metropolitan University, Osaka 599-8531, Japan}
\begin{abstract}
We study nonequilibrium steady states (NESSs) in the weakly-coupled XXZ model in contact with two heat baths at different temperatures. We show that the density matrix can be represented using only projection operators specified by the quantum numbers of the quantum group $U_q(sl_2)$ in a very good approximation. By using this property, we numerically calculate physical quantities such as temperature profile, energy current, and correlation functions, for the spin chain consisting of several hundred spins. We analytically derive the exact density matrix in the limit $q\rightarrow 0$.
\end{abstract}

\maketitle

\section{Introduction}
The understanding and description of nonequilibrium states of many-body systems is one of the most challenging problems in physics. Among various classes of nonequilibrium states, nonequilibrium steady states (NESSs) may be a promising ground for developing such theories.

In a normal heat-conducting NESS, we observe a uniform non-zero temperature gradient. However, when considering a simple theoretical model, we often observe a flat temperature profile in the system, which indicates ballistic transport. It is commonly believed that normal heat-conducting states are realized in non-integrable systems.\cite{Lepri} However, non-integrable systems can be treated almost only numerically, since its analytic treatment is quite difficult. The purpose of this paper is to find NESS solutions that are closer to normal heat-conducting states and allow more analytic treatment. In this paper, we consider the weakly-coupled XXZ model in contact with two heat baths at different temperatures. Despite the integrability of the XXZ model, numerical calculations suggest the existence of diffusive heat-conducting states.\cite{Prosen, Michel, Wu, Mendoza}

In theoretical studies of such NESSs, the coupling between the system and the bath is often assumed to be weak.\cite{Yuge, Sugita} However, if the system-bath coupling is weak and the spin-spin interaction is not weak, a flat temperature profile emerges because the thermal resistance at the boundary is large compared with that in the system. Therefore, we take the weak-coupling limit both for the spin-spin coupling in the system and for the system-bath coupling.\cite{Ishida} In this situation, the thermal resistance uniformly becomes infinitely large both in the system and at the boundary. Therefore we can expect a uniform non-zero temperature gradient in the system. We also expect that we can treat the NESS more analytically in this limit. 

For the calculation of NESSs in a quantum spin chain, tDMRG(time-dependent Density Matrix Renormalization Group) is often used.\cite{Prosen, tdmrg} However, in our weak-coupling model, obtaining the steady state using tDMRG is too time-consuming due to slow relaxation. Therefore such weakly-coupled NESSs are barely studied in the literature. We treat this problem based on the perturbative method developed in\cite{Ishida}.

We found that the density matrix for the NESS can be represented using only projection operators specified by the quantum number of the quantum group in a very good approximation, which is confirmed in terms of fidelity. By using this property, we numerically calculate the NESS simply by finding an eigenvector of a matrix whose dimension is $O(N^2)$, where $N$ is the spin number. We also calculate physical quantities such as temperature profile, energy current, and correlation functions, of the spin chain consisting of several hundred spins up to $N = 250$. In the high-temperature regime, the temperature profile shows uniform gradient. In this case, the scaling of the energy current is superdiffusive. In the low-temperature regime, the temperature profile becomes flat and the energy current consists of ballistic and diffusive parts.

This paper is organized as follows. In section 2, we briefly review the perturbative expansion for weakly-coupled steady states developed in\cite{Ishida} and its zero-th order solution. In section 3, we show that the density matrix has a simple structure that can be represented using only projection operators specified by the quantum numbers of the quantum group $U_q(sl_2)$. We also show that the weight coefficients for the projection operators can be approximated by a simple exponential form. In section 4, we analytically derive the exact density matrix in the limit $q \rightarrow 0$. In section 5, we show numerical results of the temperature profile, energy current, and correlation functions, of the spin chain consisting of several hundred spins. Section 6 is devoted to a summary.

\section{Quantum Master Equation and Its Perturbative Steady Solution}
In this section, we review the perturbative expansion for weakly-coupled NESS.\cite{Ishida} We start with the Hamiltonian
\begin{equation}
H_{total} =H_S+H_B+\lambda H_{SB}, 
\label{Hamil}
\end{equation}
where $\lambda$ is a small coupling parameter, $H_S$ is the Hamiltonian for the system, $H_B$ is that for the heat bath, and $H_{SB}$ is that for the interaction between the system and the heat bath.

Now we consider a weakly-coupled one-dimensional spin system with nearest-neighbor interactions. We assume that the spin-spin interaction is of the same order as the system-bath interaction. Then the Hamiltonian for the system can be written as 
\begin{equation}
H_S=H_0+\lambda H_I, 
\label{Hamil_S}
\end{equation}
where $H_0$ represents the external field applied to the spins and $H_I$ contains the interaction among the spins and boundary terms.

Applying the Born-Markov approximation, we obtain a Redfield-type quantum master equation (QME):
\begin{gather}
\frac{d}{dt}\rho=\mathcal{L}\rho,\label{QME} \\ 
\mathcal{L}=\mathcal{L}_0+\lambda \mathcal{L}_1+\lambda^2\mathcal{L}_2, \label{QME_L}
\end{gather}
where
\begin{align}
\mathcal{L}_0\rho&=\frac{1}{i\hbar}\lbrack H_0,\rho\rbrack, \label{QME_L0}\\
\mathcal{L}_1\rho&=\frac{1}{i\hbar}\lbrack H_1,\rho\rbrack, \label{QME_L1}\\
\mathcal{L}_2\rho&=\mathcal{D}_L\rho+\mathcal{D}_R\rho\label{QME_L2}.
\end{align}
Here,
$
H_1=H_I+\overline{H}_{SB}
$
where the overline denotes the average with respect to the heat baths.
Hence $\overline{H}_{SB}$ is an operator acting only on the system.
$\mathcal{D}_L$ and $\mathcal{D}_R$ are heat bath superoperators for the left and right heat baths, respectively. (See Appendix\ref{superoperator} 
for the heat bath superoperators in detail.) In this paper we set $\hbar =1$.

The steady state is determined by the equation
\begin{equation}
\frac{d}{dt}\rho=\mathcal{L}\rho=0. \label{steady}
\end{equation}
We calculate the steady state by expanding $\rho$ with respect to the coupling parameter $\lambda$: 
\begin{equation}
\rho=\rho_0+\lambda\rho_1+\lambda^2\rho_2 + \dots. \label{rho_pertur}
\end{equation}
By substituting it into Eq. (\ref{steady}), we obtain the following equations up to the second order:
\begin{align}
\mathcal{L}_0\rho_0=0, \label{L0_rho} &\\ 
\mathcal{L}_0\rho_1+\mathcal{L}_1\rho_0=0, \label{L1_rho} &\\
\mathcal{L}_0\rho_2+\mathcal{L}_1\rho_1+\mathcal{L}_2\rho_0=0. \label{L2_rho}
\end{align}

In usual perturbation theories, the zero-th order solution is determined from the zero-th order equation. In this case, however, we need not only Eq. (\ref{L0_rho}) but also Eqs. (\ref{L1_rho}) and (\ref{L2_rho}) to determine $\rho_0$. Physically speaking, this is because the Liouvillian $\mathcal{L}_0+\lambda\mathcal{L}_1$ represents the Hamiltonian dynamics governed by $H_0+\lambda H_1$, and all energy eigenstates of $H_0+\lambda H_1$ are stationary in this dynamics. We assume that the heat baths, which are represented by $\mathcal{L}_2$, make the stationary state unique. Then we denote the projection superoperators to $\mathrm{Ker}\mathcal{L}_0$ and $\mathrm{Ker}(\mathcal{P}_0\mathcal{L}_1\mathcal{P}_0)$ by $\mathcal{P}_0$ and $\mathcal{P}_1$, respectively. By applying the projection superoperators $\mathcal{P}_0$ and $\mathcal{P}_1$ to the equations, we obtain the following equations for $\rho_0$:
\begin{align}
\mathcal{L}_0\rho_0=0, \label{stedy_L0} &\\
\mathcal{P}_0\mathcal{L}_1\rho_0=0, \label{stedy_L1} &\\
\mathcal{P}_1\mathcal{L}_2\rho_0=0. \label{stedy_L2}
\end{align}
The zero-th order solution $\rho_0$ is determined from these equations.

In the following sections of this paper, we focus on $\rho_0$.
Note that $\rho_0$ can not be obtained just by putting $\lambda = 0$ in Eq. (\ref{steady}).
Physically speaking, $\rho_0$ is obtained by first considering a steady state at a non-zero value of $\lambda$, and 
then taking the limit $\lambda \rightarrow 0$. In this case, $\rho_0$ becomes a state with correlations, rather than just a tensor product of local equilibrium states.
It is reasonable to expect that a weak-coupling steady state can be well approximated by
$\rho_0$ for sufficiently small $\lambda$, as numerically confirmed in
\cite{Ishida} for the tilted Ising model. Although $\rho_0$ has no current,
we can readily calculate the leading order term of the energy current,
whose order is $O(\lambda^2)$, from $\rho_0$ . 
(See Section \ref{current} and Appendix \ref{Energy current} for
details.)

\section{Density Matrix Represented by Quantum Group Projection Operators}
We consider an open $N$-spin chain with the Hamiltonian 
\begin{gather}
H=H_0+\lambda H_1, \label{Hamil_XXZ}\\
H_0=h\sum_{l=1}^N\sigma_l^z, \label{Hamil_H0}\\
H_1=\sum_{l=1}^{N-1}\lparen\sigma_l^x\sigma_{l+1}^x+\sigma_l^y\sigma_{l+1}^y+\Delta\sigma_l^z\sigma_{l+1}^z\rparen+\Gamma\lparen\sigma_1^z-\sigma_N^z\rparen, \label{Hamil_H1}
\end{gather}
where $\sigma_l^\alpha (\alpha=x,y,z)$ represents the Pauli matrix at the $l$-th site, $\Delta$ is the anisotropy, and $h$ is the strength of the uniform external magnetic field. In this paper we set $h=1$. To make the system invariant under the action of the quantum group $U_q(sl_2)$, we set 
\begin{align}
\Delta&=\frac{q+q^{-1}}{2}, \label{delta} \\
\Gamma&=\frac{q-q^{-1}}{2}\label{gamma}.
\end{align}
Note that the boundary terms may depend both on $H_I$ and $\overline{H}_{SB}$. Therefore, to realize $\Gamma$ in Eq. (\ref{gamma})
in a real physical system, delicate control of the heat baths or
external field is required. However, we assume this form of the Hamiltonian just for theoretical simplicity.
In this paper we assume that $q$ is a real number to make the Hamiltonian Eq. (\ref{Hamil_XXZ}) hermitian. Note that the replacement of $q$ with $q^{-1}$ corresponds to the spatial inversion of the system. In this paper we consider $0<q<1$ for numerical calculations. 

We assume that the left (right) heat bath is in equilibrium with the inverse temperature $\beta_L$ ($\beta_R$). We also define 
\begin{equation}
\bar{\beta}=\frac{\beta_{L}+\beta_{R}}{2},
\label{beta_bar}
\end{equation}
and
\begin{equation}
\Delta\beta=\beta_{L}-\beta_{R}. 
\label{beta_delta}
\end{equation}
In our weak-coupling model, a heat bath superoperator acts only on a edge spin. Its explicit form is given in Appendix\ref{superoperator}.

\subsection{Structure of Steady State Density Matrix}
In the XXZ model Eq. (\ref{Hamil_XXZ}), $H_0$ and $H_1$ can be diagonalized simultaneously. Then Eq. (\ref{stedy_L1}) means that $\rho_0$ is diagonal in the energy representation:
\begin{equation}
\rho = \sum_{i=1}^{2^N} C_i \ket{E_i}\bra{E_i}, \label{rho_energy}
\end{equation}
where $\ket{E_i}$ is the eigenvector of $H$ corresponding to the $i$-th eigenvalue $E_i$ and $C_i$ is the weight coefficient for the state.
Note that  
the eigenvector $\ket{E_i}$ for the Hamiltonian (\ref{Hamil_XXZ}) does not depend on
the coupling parameter $\lambda$, while the eigenvalue $E_i$ depends
on $\lambda$.

\begin{figure}[H]
\begin{tabular}{cc}
\begin{minipage}{.5\textwidth}
\centering
\includegraphics[width=1\linewidth]{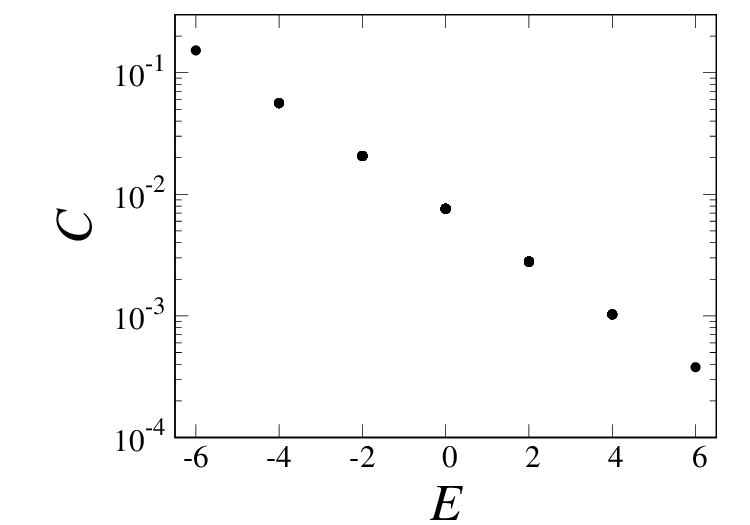}
\end{minipage}
\begin{minipage}{.5\textwidth}
\centering
\includegraphics[width=1.0\linewidth]{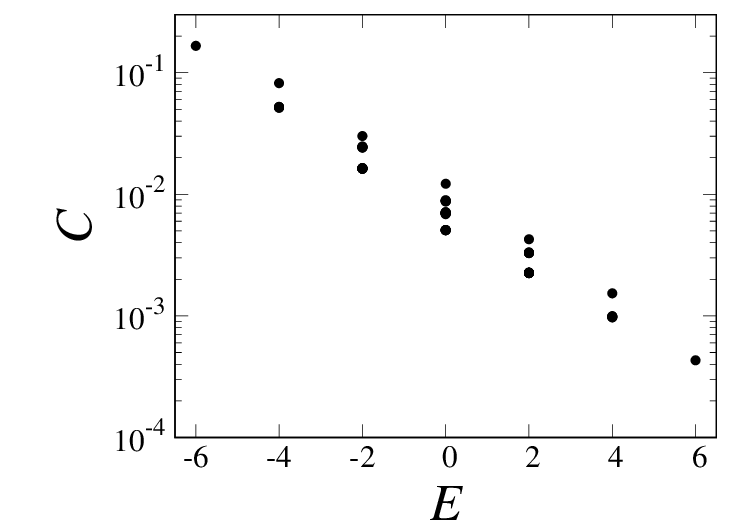}
\end{minipage}
\end{tabular}
\caption{Numerical results of the weight coefficients for our model with $N=6$, $\bar{\beta}=0.5$, and $q=0.5$. 
The horizontal axis shows the eigenenergies of the states at $\lambda = 0$. The left and right figures correspond to $\Delta{\beta}=0$ (equilibrium state) and $\Delta{\beta}=0.3$ (nonequilibrium steady state), respectively. Although there are $2^6=64$ points in each figure, we see fewer points because of the degeneracy.}
\label{weight}
\end{figure}

We show that the density matrix for the steady state has a simpler structure that can be represented using only projection operators specified by the quantum numbers of the quantum group. Fig. \ref{weight} shows numerical results of the weight coefficients. In the left figure, we observe a linear shape in the logarithmic plot because, in an equilibrium state, $C_i$ is determined by
\begin{equation}
C_i=\frac{1}{Z}\exp(-\beta E_i), \label{weight_coeffi}
\end{equation}
where $\beta$ is the inverse temperature and $Z$ is the partition function. Note that the energy eigenvalues of $H$ are determined only by the $z$-component of angular momentum in the limit $\lambda\rightarrow 0$. Therefore the energy eigenvalues are highly degenerate, and the weight coefficients are also highly degenerate in the equilibrium case.

In the right figure which represents the nonequilibrium case, we see that the degeneracy is partially solved but some points remain almost degenerate. For example, although there should be 20 points at $E=0$, we see only 4 points. The eigenvectors corresponding to the degenerate weight coefficients are in the same irreducible representation of the quantum group $U_q(sl_2)$. Hence, the steady states can be represented only by the projection operators corresponding to the quantum number of the quantum group: 
\begin{gather}
\rho=\sum_{J,m}C_{J,m}P_{J,m}, \label{rho_J,m} \\
P_{J,m}=\sum_i\ket{J,m,i}\bra{J,m,i}, \label{P_J,m}
\end{gather}
where $\ket{J,m,i}$ represents a state characterized by the quantum numbers $J$ and $m$, $J$ is the quantum number to specify the irreducible representation, and $m$ is the angular momentum in the $z$-direction. Note that $J$ is similar to, but not the same as, the total angular momentum. $J$ and $m$ are integers or half-integers depending on the parity of the spin number $N$, which satisfy $0\le J\le N/2$ and $-J\le m \le J$.

\subsection{Matrix Representation of the Heat Bath Superoperator by the Quantum Group Projection Operators}
\label{Matrix repre}
The projection operator Eq. (\ref{P_J,m}) can be constructed algebraically. (See Appendix\ref{projection operators}.) The weight coefficient $C_{J,m}$ in Eq. (\ref{rho_J,m}) is determined from Eq. (\ref{stedy_L2}), which is accomplished by expressing the heat bath superoperators $\mathcal{D}_L$ and $\mathcal{D}_R$ in matrix form and subsequently finding the eigenvector of $\mathcal{D}_L + \mathcal{D}_R$ belonging to the eigenvalue zero.

If we do not introduce any approximation, the number of the weight coefficients in Eq. (\ref{rho_energy}), which determines the matrix dimension of the heat bath superoperators, is $2^N$. However, in the projection operator representation Eq. (\ref{rho_J,m}), the number of the weight coefficients is only $O(N^2)$, which dramatically reduces the amount of computation. See Appendix\ref{D_element} for the calculation of matrix elements in detail.

\subsection{Fidelity}
\begin{figure}[H]
\begin{center}
\includegraphics[width=100mm]{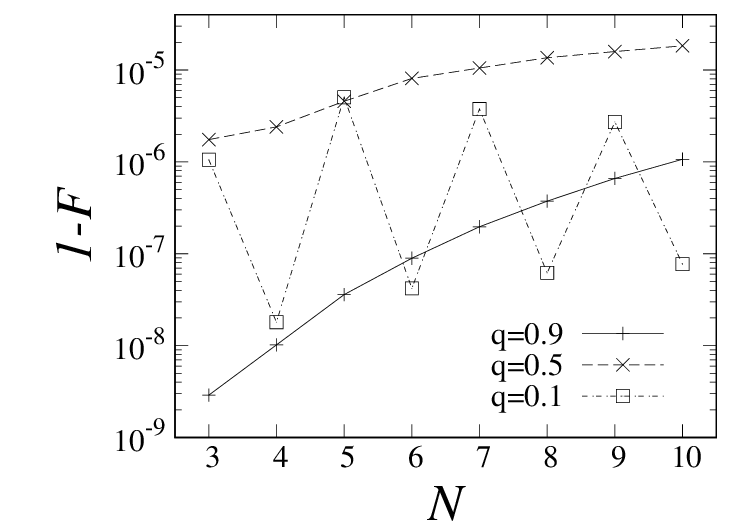}
\caption{Numerical results of the fidelity loss $1-F$ for the projection operator representation of the density matrix up to $N=10$ with $\bar{\beta}=0.5$ and $\Delta{\beta}=0.3$.}
\label{fidelity}
\end{center}
\end{figure}

We demonstrate that the density matrix represented by the quantum group projection operators closely approximates the exact density matrix. We use fidelity  
\begin{equation}
F\lparen\rho_1,\rho_2\rparen=
\left(\operatorname{Tr}\sqrt{\sqrt{\rho_2}\rho_1\sqrt{\rho_2}}\right)^{2}, \label{fideli}
\end{equation}
to quantify the closeness of two quantum states. Fig. \ref{fidelity} shows numerical result of the fidelity loss $1-F(\rho_1, \rho_2)$, where $\rho_1$ is numerically obtained steady state density matrix without approximation, and $\rho_2$ is that obtained assuming the projection operator representation Eq. (\ref{rho_J,m}). We see that the fidelity loss is of the order $10^{-8}$ to $10^{-5}$, which increases with $N$ but not rapidly. In the case with $q=0.1$, we see that the fidelity loss depends on the parity of $N$ and is larger for odd $N$. However, it remains about $10^{-6}$, which still can be considered small. Therefore the projection operator representation Eq. (\ref{rho_J,m}) can be considered a good approximation to the exact steady state.

\subsection{Weight Coefficients}
\label{wei_coeff}
\begin{figure}[H]
\begin{center}
\includegraphics[width=150mm]{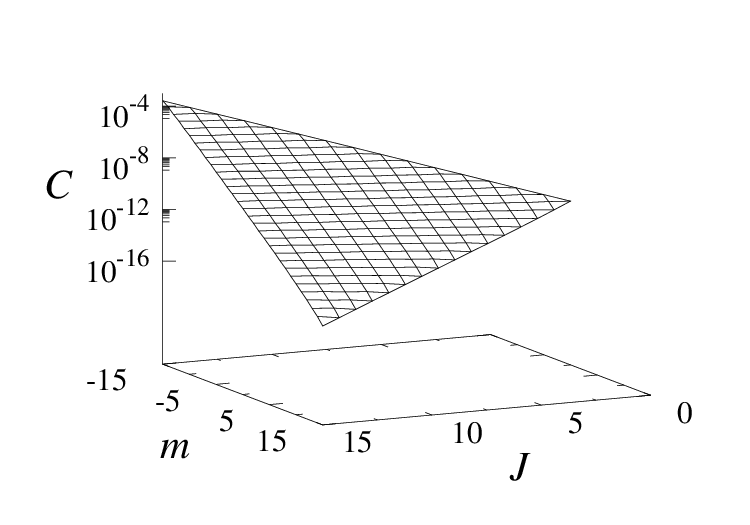}
\caption{Weight coefficients for the projection operators with $N=30$, $\bar{\beta}=0.5$, $\Delta{\beta}=0.3$, and $q=0.5$.}
\label{3jigen}
\end{center}
\end{figure}

In Fig. \ref{3jigen} we show an example of the weight coefficients. The logarithmic graph looks like an almost flat plane, which suggests that $\log C_{J,m}$ is a linear function of $J$ and $m$ approximately. Thus we have
\begin{equation}
C_{J,m} \propto \exp(\gamma J + \delta m), \label{weigh analy}
\end{equation}
where $\gamma$ and $\delta$ are some constants.

Fig. \ref{every J,m q=0.5} and Fig. \ref{every J,m q=0.1} show the weight coefficients in the case with $J=10$ and $m=1$, respectively for $q=0.5$ and $q=0.1$. We can confirm that the shapes of the graph are almost linear, but for $q=0.1$, we should introduce the parity of $J$ to represent the weight coefficients more precisely. If we treat even $J$ and odd $J$ separately, we see that the shapes of the graph are also linear.  A closer inspection shows that the parity dependence also exists for $m$.
 
\begin{figure}[H]
\begin{tabular}{cc}
\begin{minipage}{.5\textwidth}
\centering
\includegraphics[width=1.0\linewidth]{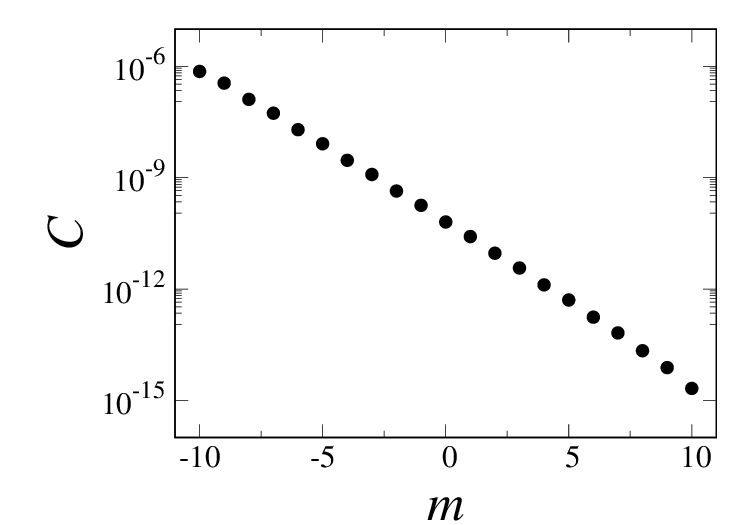}
\end{minipage}
\begin{minipage}{.5\textwidth}
\centering
\includegraphics[width=1\linewidth]{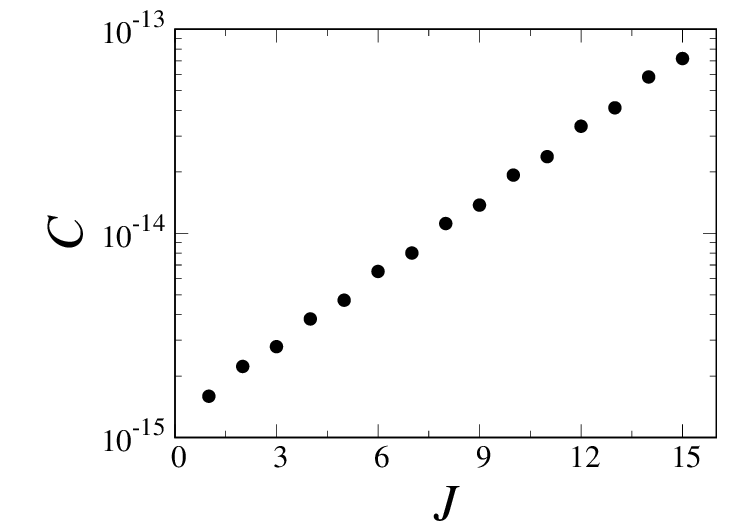}
\end{minipage}
\end{tabular}
\caption{Weight coefficients for the projection operators with fixed values of $J$ and $m$ for $N=30$ and $q=0.5$. The left and right figures correspond to $J=10$ and $m=1$, respectively. The other parameters, which are not written explicitly here, are the same as those in Fig. \ref{3jigen}.}
\label{every J,m q=0.5}
\end{figure}

\begin{figure}[H]
\begin{tabular}{cc}
\begin{minipage}{.5\textwidth}
\centering
\includegraphics[width=1.0\linewidth]{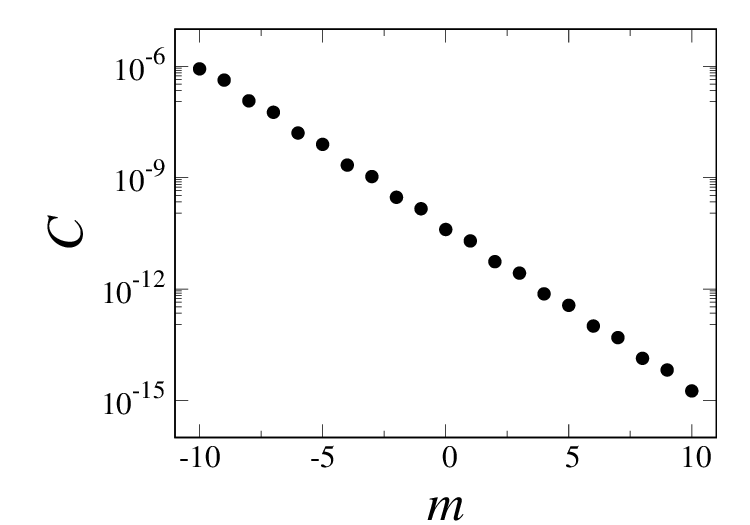}
\end{minipage}
\begin{minipage}{.5\textwidth}
\centering
\includegraphics[width=1\linewidth]{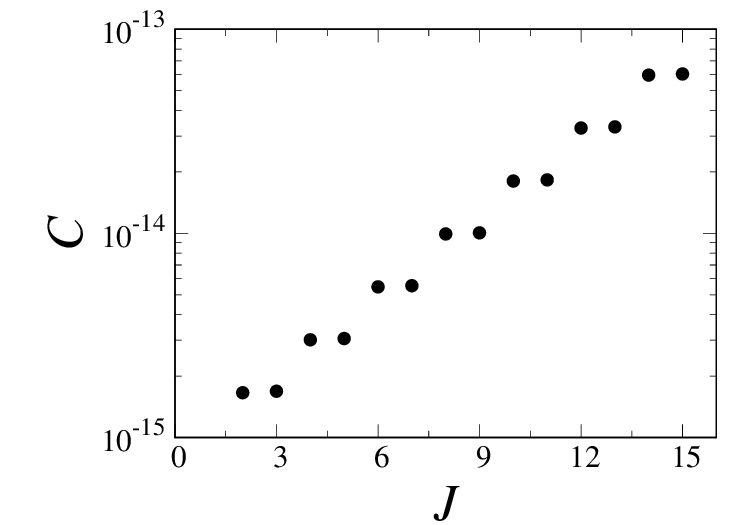}
\end{minipage}
\end{tabular}
\caption{Weight coefficients for q=0.1. The other parameters, which are not written explicitly here, are the same as those in Fig. \ref{every J,m q=0.5}.}
\label{every J,m q=0.1}
\end{figure}

\section{Analytic Solution in the Limit $q\rightarrow 0$}
In this section, we analytically derive the exact density matrix for the zero-th order steady state represented by the quantum group projection operators in the limit $q\rightarrow 0$ for even $N$. 

Let us first consider recurrence relations for the projection operator by using the Clebsch-Gordan coefficients for the quantum group. (See Appendix\ref{projection operators} for details.) For example, if $J$ and $m$ are not at the edges, i.e. $m\ne \pm J$ and $J\ne 0$, the recurrence relations for the projection operators are
\begin{gather}
\sigma_u\otimes P_{N,J,m}\otimes \sigma_u=P_{N+2,J,m+1}, \label{up_P_up} \\
\sigma_d\otimes P_{N,J,m}\otimes \sigma_u=P_{N+2,J+1,m}, \label{down_P_up} \\
\sigma_u\otimes P_{N,J,m}\otimes \sigma_d=P_{N+2,J-1,m}, \label{up_P_down} \\
\sigma_d\otimes P_{N,J,m}\otimes \sigma_d=P_{N+2,J,m-1}, \label{down_P_down}
\end{gather}
where $P_{N,J,m}$ denotes the projection operator for $N$-spin system with the quantum numbers $J$ and $m$,  $\sigma_u$ and $\sigma_d$ are the density matrices for up spin and down spin:
\begin{gather}
\sigma_{u}=\ket{\uparrow}\bra{\uparrow}, \label{sigma_u} \\
\sigma_{d}=\ket{\downarrow}\bra{\downarrow}. \label{sigma_d}
\end{gather}
When the heat baths, whose explicit form is given in Appendix\ref{superoperator}, act on the $(N+2)$-spin system, the weight coefficients satisfy the following equations:
\begin{eqnarray}
\begin{pmatrix}
t_{L}+t_{R}+2 & t_{L}-1 & t_{R}-1 & 0 \\
-t_{L}-1 & -t_{L}+t_{R}+2 & 0 & t_{R}-1 \\
-t_{R}-1 & 0 & t_{L}-t_{R}+2 &  t_{L}-1 \\
0 & -t_{R}-1 & -t_{L}-1 & -t_{L}-t_{R}+2  \\
\end{pmatrix}
\begin{pmatrix}
C_{J,m+1}  \\
C_{J+1,m}  \\
C_{J-1,m}  \\
C_{J,m-1}  \\
\end{pmatrix}
=0,
\label{rho_t=0}
\end{eqnarray}
where $t_l=\tanh(\beta_l h)$ $(l=L,R)$ is determined by the temperatures of the heat baths. The recurrence relations for $m=\pm J$ and $J=0$ can be obtained in the same way:
\begin{eqnarray}
\mathcal{D}
\begin{pmatrix}
C_{J+1,J+1}  \\
C_{J+1,J}  \\
C_{J,J}  \\
C_{J,J-1}  \\
\end{pmatrix}
=
\mathcal{D}
\begin{pmatrix}
C_{J,-J+1}  \\
C_{J+1,-J}  \\
C_{J,-J}  \\
C_{J+1,-J-1}  \\
\end{pmatrix}
=
\mathcal{D}
\begin{pmatrix}
C_{1,1}  \\
C_{1,0}  \\
C_{0,0}  \\
C_{1,-1}  \\
\end{pmatrix}
=0,
\label{rho=0}
\end{eqnarray}
where $\mathcal{D}$ is the matrix in Eq. (\ref{rho_t=0}). 

From these equations we obtain the explicit form for the weight coefficients
\begin{equation}
C_{J,m}\propto (1-t_L)^{\frac{N}{2}-[\frac{J-m+1}{2}]}(1+t_L)^{[\frac{J-m+1}{2}]}(1-t_R)^{J-[\frac{J-m}{2}]}(1+t_R)^{\frac{N}{2}-J+[\frac{J-m}{2}]}, 
\label{weigh analy_q0_t}
\end{equation}
where $[\cdot]$ denotes the floor function. This can be rewritten as
\begin{equation}
C_{J,m} \propto \exp\bigg\{-2\bar{\beta}hm+\Delta\beta h
\left(J-\frac{(-1)^{J+m}}{2}\right)\bigg\}.    
\label{weight rewrite}
\end{equation}
We see that the index part is a linear function with respect to $J$ and $m$ accompanied by a parity correction term. This is in accordance with the observation in Section \ref{wei_coeff} based on the numerical results.

For odd $N$, an exact steady state solution is not obtained if we assume the projection operator representation of the density matrix Eq. (\ref{rho_J,m}). This observation may suggest that this approximation is inadequate for odd $N$ and $q\rightarrow 0$.

\section{Physical Quantities}
In this section, we show the results of numerical calculations based on the method explained in Section \ref{Matrix repre}. We calculated up to about $N=250$. For the 250 spins, the dimension of the matrix representation of the heat bath superoperator
is $(N+2)^2/4=15876$.

\subsection{Temperature Profile}
In our model, $\braket{\sigma_n^x}$ and $\braket{\sigma_n^y}$ vanish because of the symmetry of the system. Then the reduced density matrix $\rho_n$ for the $n$-th site is represented by $\sigma_n^0$ and $\sigma_n^z$:
\begin{equation}
\rho_n = \frac{1}{2}\left(\sigma_n^0 +  \braket{\sigma_n^z}\sigma_n^z\right). \label{rho_n}
\end{equation}
This can be identified with an equilibrium state with respect to the single site Hamiltonian $h\sigma_n^z$
\begin{equation}
\frac{1}{Z}\exp(-\beta_n h\sigma_n^z )=\frac{1}{2}\big(\sigma_n^0-\tanh(\beta_n h)\sigma_n^z\big).
\label{Hamil_single}
\end{equation}
Thus we can define the local inverse temperature $\beta_n$ by the following equation:
\begin{equation}
\braket{\sigma_n^z}=-\tanh(\beta_n h). \label{local temp}
\end{equation}
Numerically we can calculate $\braket{\sigma_n^z}$ for the steady state easily if we know the expectation values of $\sigma_n^z$ for the projection operators $P_{J,m}$. They are obtained from recurrence relations in Appendix\ref{recursion}.

\begin{figure}[H]
\begin{tabular}{cc}
\begin{minipage}{.5\textwidth}
\centering
\includegraphics[width=1.0\linewidth]{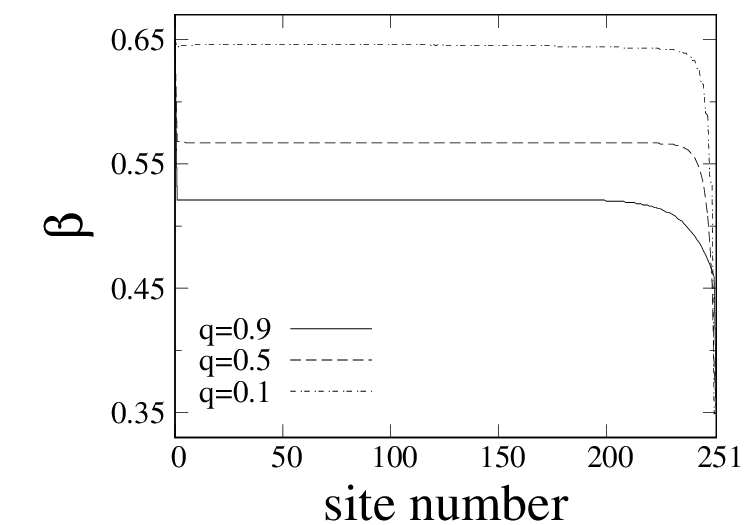}
\end{minipage}
\begin{minipage}{.5\textwidth}
\centering
\includegraphics[width=1\linewidth]{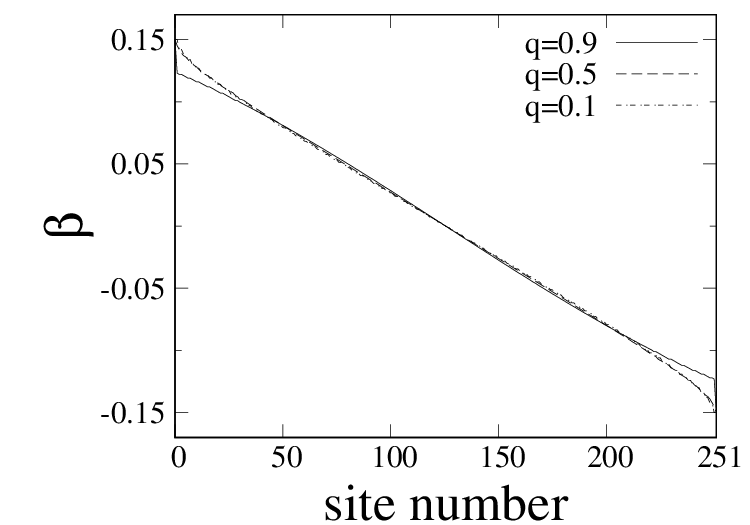}
\end{minipage}
\end{tabular}
\caption{Temperature profiles for $N=250$ and $\Delta{\beta}=0.3$. The left and right figures correspond to $\bar{\beta}=0.5$ and $\bar{\beta}=0$, respectively. The heat bath inverse temperatures are shown at the 0th and 251th sites in the figure, respectively.} 
\label{temp q}
\end{figure}

\begin{figure}[H]
\begin{center}
\includegraphics[width=100mm]{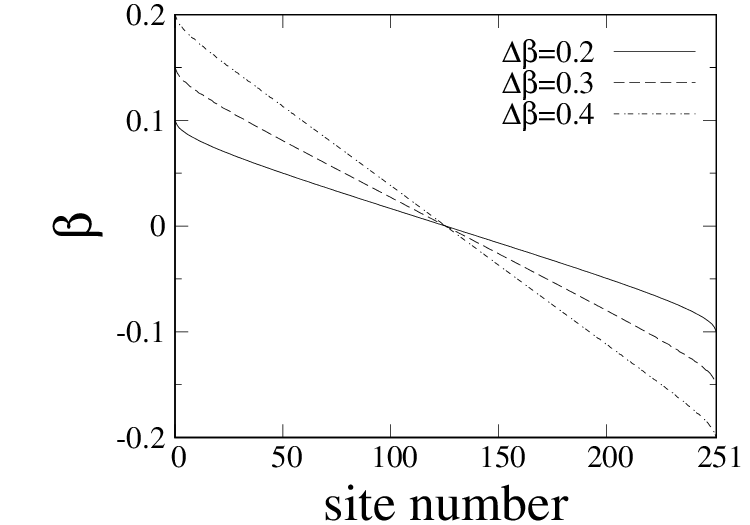}
\caption{Temperature profiles for the various $\Delta{\beta}$ with $N=250$, $\bar{\beta}=0$, and $q=0.5$.}
\label{temp dt}
\end{center}
\end{figure}

Fig. \ref{temp q} and Fig. \ref{temp dt} show the temperature profiles for our model with $N$ = 250. For $\bar{\beta}=0.5$, the temperature gradient is almost flat except near the boundaries. In the symmetrically driven case $\bar{\beta}=0$, which represents the high-temperature regime, a uniform temperature gradient is formed.

\begin{figure}[H]
\begin{tabular}{cc}
\begin{minipage}{.5\textwidth}
\centering
\includegraphics[width=1.0\linewidth]{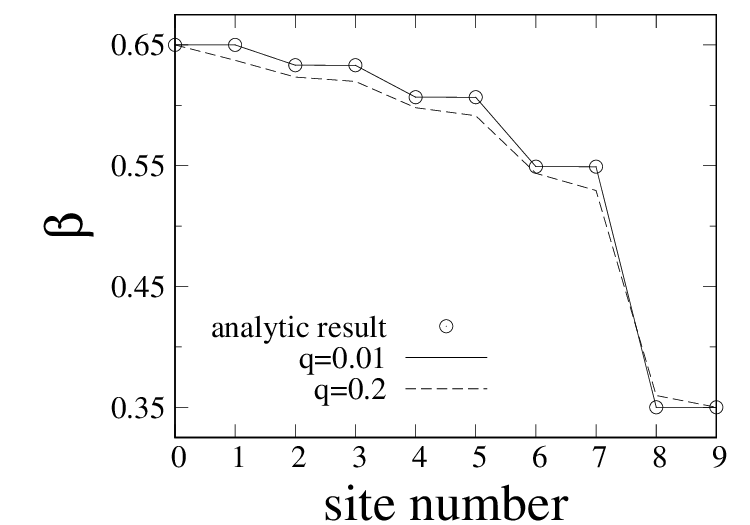}
\end{minipage}
\begin{minipage}{.5\textwidth}
\centering
\includegraphics[width=1\linewidth]{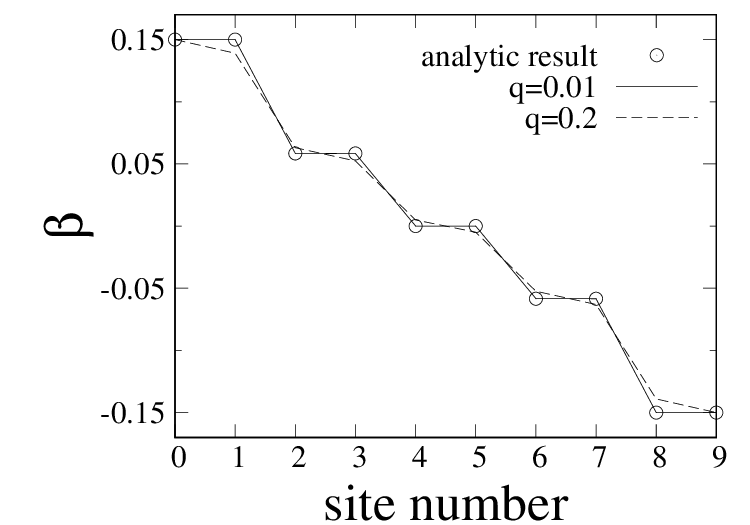}
\end{minipage}
\end{tabular}
\caption{Comparison of temperature profile obtained
from the analytic solution for $q\rightarrow 0$ and that obtained from numerical calculation with $N=8$ and $\Delta{\beta}=0.3$. The left and right figures correspond to $\bar{\beta}=0.5$ and $\bar{\beta}=0$, respectively. The heat bath inverse temperatures are shown at the 0th and 9th sites in the figure, respectively. In the case with $q=0.01$, the numerical results are almost the same as the analytic ones.}
\label{temp analytic}
\end{figure}

Fig. \ref{temp analytic} shows a comparison between numerical results and analytic ones for $q\rightarrow 0$ based on Eq. (\ref{weight rewrite}). We see that analytic results and numerical ones for $q=0.01$ almost coincide. The results for $q=0.2$ are also not so far from the analytic ones.

\subsection{Energy Current}\label{current}
Here we consider the energy current. In our model, time derivative of the energy of the system is given by
\begin{equation}
\frac{d}{dt}\braket{H_S}=\lambda^2 \operatorname{Tr}(\lparen \mathcal{D}_L\rho\rparen H_S)+ \lambda^2\operatorname{Tr}(\lparen \mathcal{D}_R\rho\rparen H_S), \label{expectation_Hs}
\end{equation}
where $\mathcal{D}_L$ and $\mathcal{D}_R$ are superoperators for the heat baths. The two terms can be regarded as the energies coming from the two heat baths. In the weak-coupling  case, the energy current is $O(\lambda^2)$,
and the leading order term is easily derived from the zero-th order solution
$\rho_0$.
We define the scaled energy current for the steady state as
\begin{equation}
j=\operatorname{Tr}(\lparen{\mathcal{D}_L\rho}\rparen H_S).
\label{j before}
\end{equation}
For our model the energy current is explicitly written as
\begin{equation}
j=ah\lparen t_1-t_L\rparen,
\label{j}
\end{equation}
where $a$ is a parameter for the heat bath explained in Appendix\ref{superoperator} and $t_l = \tanh(\beta_l h)$. Therefore we can calculate the energy current $j$ easily from the temperature of the edge spin. See Appendix\ref{Energy current} for the derivation of Eq. (\ref{j}).

\begin{figure}[H]
\begin{tabular}{cc}
\begin{minipage}{.5\textwidth}
\centering
\includegraphics[width=1.0\linewidth]{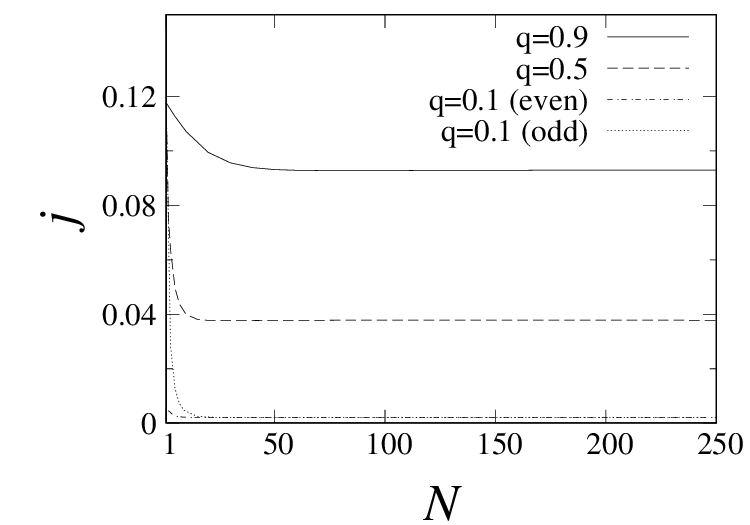}
\end{minipage}
\begin{minipage}{.5\textwidth}
\centering
\includegraphics[width=1.0\linewidth]{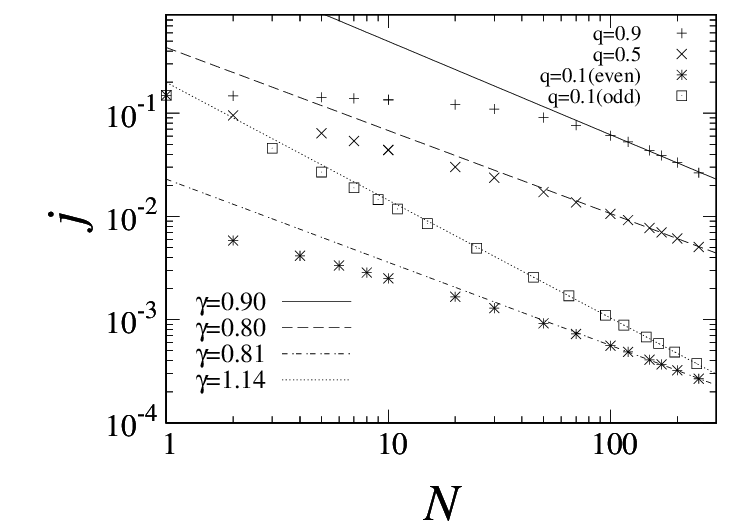}
\end{minipage}
\end{tabular}
\caption{Energy current up to $N=250$ with $\Delta{\beta}=0.3$. The left and right figures correspond to $\bar{\beta}=0.5$ and $\bar{\beta}=0$, respectively. The right is a log-log plot. For $q=0.1$ the data are plotted separately for even and odd $N$.} 
\label{energy current}
\end{figure}

\begin{figure}[H]
\begin{tabular}{cc}
\begin{minipage}{.5\textwidth}
\centering
\includegraphics[width=1.0\linewidth]{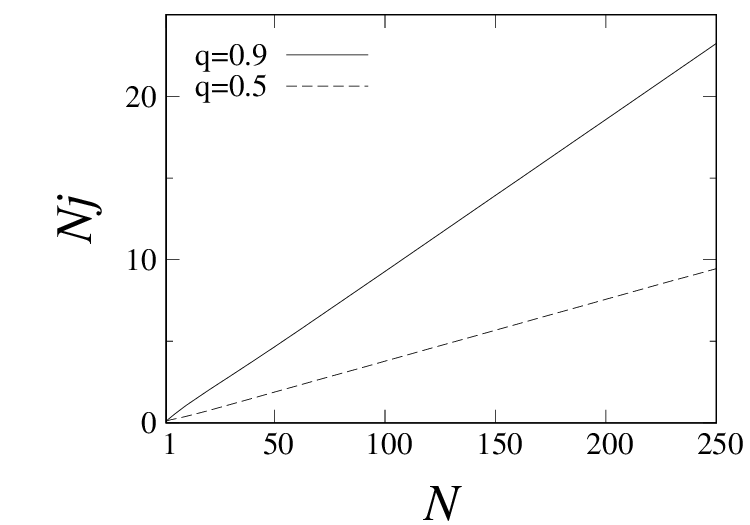}
\end{minipage}
\begin{minipage}{.5\textwidth}
\centering
\includegraphics[width=1.0\linewidth]{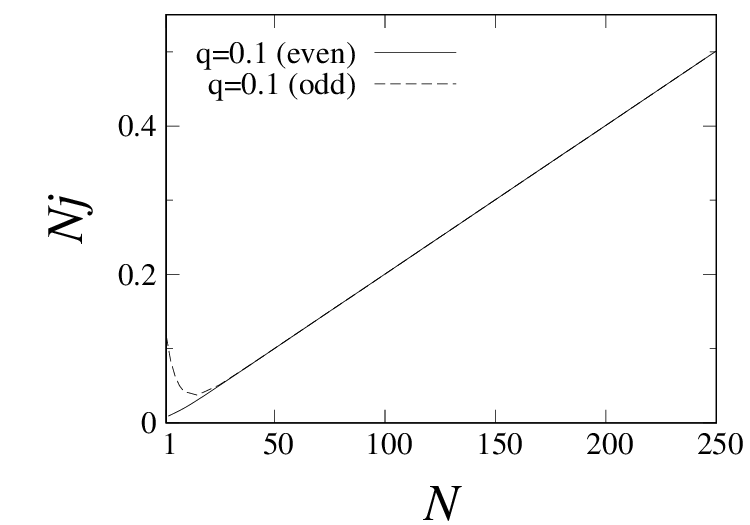}
\end{minipage}
\end{tabular}
\caption{$Nj$ up to $N=250$, $\bar{\beta}=0.5$ and $\Delta{\beta}=0.3$. The left figure shows the cases with $q=0.9$ and $q=0.5$. The right figure shows the case with $q=0.1$. In this case the data are plotted separately for even and odd $N$. $Nj$ usually becomes a linear function of $N$, but an anomalous behavior is observed for the case with $q=0.1$ and odd $N$.}
\label{NJ}
\end{figure}

Fig. \ref{energy current} shows the energy current for our model up to $N=250$ for a fixed $\Delta \beta$. In the case with $\bar{\beta}=0.5$, the energy current can be approximated by the sum of a diffusive term and a ballistic term
\begin{equation}
j=\frac{d}{N}+e, 
\label{j_diff_balli}
\end{equation}
where $d$ and $e$ are some constants. This is confirmed in Fig. \ref{NJ} which shows that $N\times j$ is a linear function of $N$. The case with small $q$ ($q=0.1$) and odd $N$ is exceptional and the energy current deviates from the approximation Eq. (\ref{j_diff_balli}) remarkably for small $N$.

In the case with $\bar{\beta}=0$, we see that $j$ is represented by 
\begin{equation}
j\propto\frac{1}{N^\gamma}, 
\label{j_superdiff}
\end{equation}
where $\gamma$ is a scaling exponent. In the case with $q=0.9$, $q=0.5$, and $q=0.1$ (even $N$) the scaling of the energy current is superdiffusive ($0<\gamma < 1$). In the case with $q=0.1$ (odd $N$) the energy current is subdiffusive ($\gamma > 1$). 

\subsection{Correlation Function}
Here we consider the two-spin correlation function 
\begin{equation}
\braket{\Delta\sigma_l^\alpha \Delta\sigma_r^\beta}=\braket{\sigma_l^\alpha \sigma_r^\beta}-\braket{\sigma_l^\alpha}\braket{\sigma_r^\beta}. 
\label{corr}
\end{equation}
We denote the distance between two spins as $d=|r-l|$. In our model, it satisfies
\begin{equation}
\braket{\Delta\sigma_l^\alpha \Delta\sigma_r^\beta}=0,
\label{corr_0}
\end{equation}
for $\alpha \ne \beta$ and
\begin{equation}
\braket{\Delta\sigma_l^x\Delta\sigma_r^x}=\braket{\Delta\sigma_l^y \Delta\sigma_r^y},
\label{corr_symm}
\end{equation}
because of the symmetry of the model.

Numerically we can calculate $\braket{\Delta\sigma_l^\alpha \Delta\sigma_r^\beta}$ for the steady state easily if we know the expectation values of $\Delta\sigma_l^\alpha \Delta\sigma_r^\beta$ for the projection operators $P_{J,m}$. They are obtained from recurrence relations in Appendix\ref{recursion}.

\begin{figure}[H]
\begin{center}
\includegraphics[width=100mm]{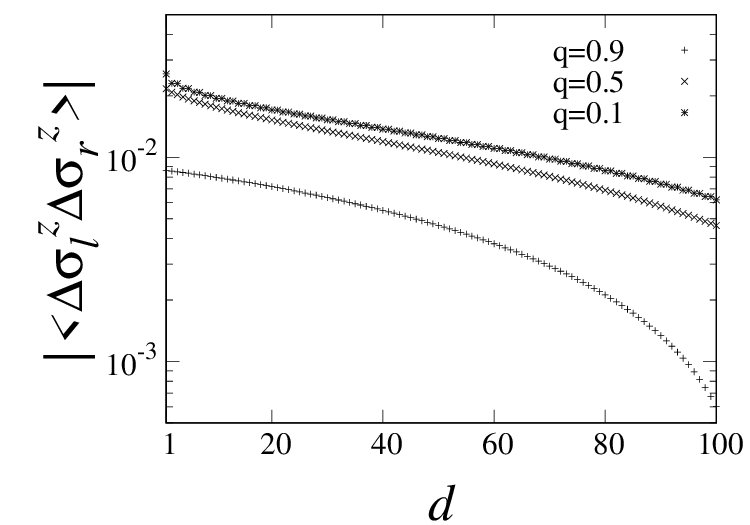}
\caption{Semi-log plot for the magnitude of the $z$-$z$ spin correlation function with $N=180$, $\Delta{\beta}=0.3$, and $\bar{\beta}=0$. We set $l=40$ and took $r$ from 41 to 140.} 
\label{zz correlation}
\end{center}
\end{figure}

Fig. \ref{zz correlation} shows the magnitude of the $z$-$z$ spin correlation function. We see that in the case with $q=0.1$ and $q=0.5$ the spin correlation function decays almost exponentially, but in the case with $q=0.9$ it deviates from the exponential law. 

\begin{figure}[H]
\begin{center}
\includegraphics[width=100mm]{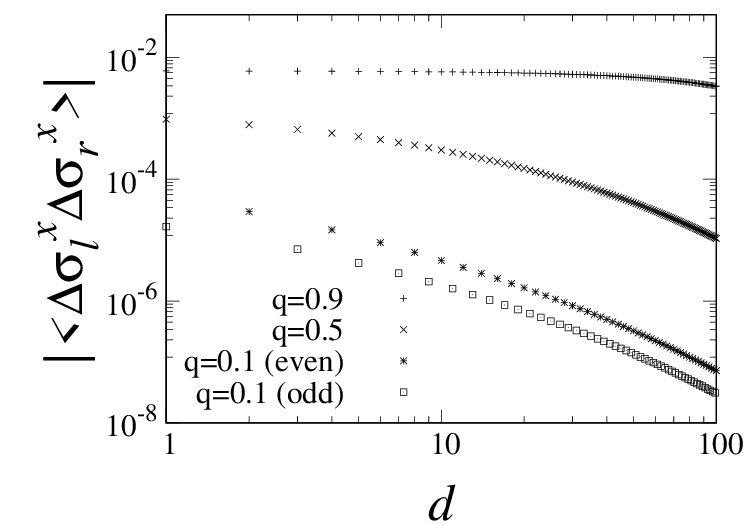}
\caption{Log-log plot for the magnitude of the $x$-$x$ spin correlation function with $N=180$. The other parameters, which are not written explicitly here, are the same as those in Fig. \ref{zz correlation}. For $q=0.1$ the data are plotted separately for even and odd $d$}
\label{xx correlation}
\end{center}
\end{figure}

Fig. \ref{xx correlation} shows the magnitude of the $x$-$x$ spin correlation function. In this case the correlation function seems to decay according to a power-law. More extensive study on the correlation functions will be reported elsewhere.

\section{Summary}
We considered NESSs in the weakly-coupled XXZ model. The spin-spin coupling and the system-bath coupling were assumed to be of the same order. We showed that the density matrix in our model has a simple structure that can be represented using only projection operators specified by the quantum number of the quantum group
\begin{equation}
\rho = \sum_{J,m}C_{J,m}P_{J,m} \label{rho_J,m after}.
\end{equation}
We confirmed the accuracy of this approximation using the fidelity. We also found that the weight coefficients are approximated by a simple exponential function
\begin{equation}
C_{J,m} \propto \exp(\gamma J + \delta m). \label{weigh analy after}
\end{equation}
By using the projection operator representation Eq. (\ref{rho_J,m after}), we numerically calculated physical quantities such as temperature profile, energy current, and correlation functions, for the spin chain consisting of several hundred spins. We also analytically derived the exact density matrix in the limit $q \rightarrow 0$ for even spin numbers. 

\begin{acknowledgments}
The authors thank Akira Terai for helpful discussions and comments.
\end{acknowledgments}

\appendix
\section{Heat Bath Superoperator}
\label{superoperator}
In this appendix we explain the heat bath superoperators in our model. We introduce the system-bath interaction
\begin{equation}
H_{SB}=\sum_j\sigma_j\otimes Y_j, \label{Hamil_I}
\end{equation}
where $\sigma_j$'s act on the edge spin of the system and $Y_j$'s act on the heat bath. Then the heat bath superoperator is
\begin{gather}
\mathcal{D}\rho=-\sum_j
\lbrack \sigma_j,S_j\rho - \rho S_j^\dagger\rbrack, \label{Drho}\\
S_j=a_{jx}\sigma_x+a_{jy}\sigma_y+a_{jz}\sigma_z, \label{S_j}
\end{gather}
where 
\begin{gather}
a_{jx}=\frac{1}{2}\{\Xi_{jx}(2h)+\Xi_{jx}(-2h)\}-\frac{i}{2}\{\Xi_{jy}(2h)-\Xi_{jy}(-2h)\}, \label{a_jx}\\
a_{jy}=\frac{1}{2}\{\Xi_{jy}(2h)+\Xi_{jy}(-2h)\}+\frac{i}{2}\{\Xi_{jx}(2h)-\Xi_{jx}(-2h)\}, \label{a_jy}\\
a_{jz}=\Xi_{jz}(0), \label{a_jz}
\end{gather}
\begin{gather}
\Xi_{jl}(\omega)=\int_{0}^{\infty}dt\,
e^{-i\omega t}\Phi_{jl}(t), \label{Xi}\\
\Phi_{jl}(t)=\braket{\Delta Y_j(t)\Delta Y_l}. \label{Phi}
\end{gather}
We replace $\Xi_{jl}$ by $\frac{1}{2}\tilde{\Phi_{jl}}$, where $\tilde{\Phi}_{jl}$ is the Fourier transform of $\Phi$:\cite{Saito}
\begin{align}
\tilde{\psi}_{jl}(\omega) = \int_{-\infty}^\infty dt\,
e^{-i\omega t} dt\Psi_{jl}(t). \label{Psi}
\end{align}
We assume that $\tilde{\Phi}_{xy}(\omega), \tilde{\Phi}_{xz}(\omega), \tilde{\Phi}_{yx}(\omega), \tilde{\Phi}_{yz}(\omega), \tilde{\Phi}_{zx}(\omega)$, and $\tilde{\Phi}_{zy}(\omega)$ vanish and $\tilde{\Phi}_{xx}(\omega) = \tilde{\Phi}_{yy}(\omega)$, which is in accordance with the symmetries of the XXZ system Hamiltonian. Then we obtain the following form of the heat bath superoperator:
\begin{align}
\mathcal{D}\sigma_0&=ta\sigma_z, \label{Dsigma_0}\\
\mathcal{D}\sigma_x&=b\sigma_x, \label{Dsigma_x}\\
\mathcal{D}\sigma_y&=b\sigma_y, \label{Dsigma_y}\\
\mathcal{D}\sigma_z&=a\sigma_z, \label{Dsigma_z} 
\end{align}
where $t=\tanh(\beta h), a=-2(1+e^{2\beta h})\tilde{\Phi}_{xx}(2h)$, and $b=\frac{a}{2}-2\tilde{\Phi}_{zz}(0)$. The choice of the parameters $a$ and $b$ does not strongly affect qualitative properties of the steady state. In this paper, we set $a=-1$, $b=-1/2$, which makes the equations simpler.

The matrix elements of the heat bath superoperator $\mathcal{D}$ in the energy representation are
\begin{equation}
{\rm Tr}(\ket{E_n}\bra{E_n}\mathcal{D}\ket{E_m}\bra{E_m})=2\sum_j\Re \lparen\bra{E_n}S_j\ket{E_m}\bra{E_m}\sigma_j\ket{E_n}-\delta_{nm}\bra{E_n}\sigma_jS_j\ket{E_m}\rparen, \label{Delement}
\end{equation}
where
\begin{align}
S_x&=-\frac{a}{8}\lparen\sigma_x-it\sigma_y\rparen, \label{S_x}\\
S_y&=-\frac{a}{8}\lparen it\sigma_x+\sigma_y\rparen, \label{S_y}\\
S_z&=\frac{a-2b}{8}\sigma_z. \label{S_z}
\end{align}

\section{Construction of the Projection Operators}
\label{projection operators}
\subsection{Clebsch-Gordan Coefficients for the Quantum Group}
Here we explain how to construct the projection operator corresponding to the quantum number $J$ and $m$ of the quantum group. The quantum group $U_q(sl_2)$ is an algebra with three generators $X_+, X_-, J_z$. For a single spin, they can be written simply as
\begin{equation}
X_+ = \begin{pmatrix}
0 & 1\\
0 & 0
\end{pmatrix},\;\;\;
X_- = \begin{pmatrix}
0 & 0\\
1 & 0
\end{pmatrix},\;\;\;
J_z = \frac{1}{2}\begin{pmatrix}
1 & 0\\
0 & -1
\end{pmatrix}.
\label{generator}
\end{equation}
For many-spin systems, the generators can be constructed by using the composition rule
\begin{align}
X_{\pm} &=X_{1,\pm}\otimes k_2^{-1}+k_1\otimes X_{2,\pm}, \label{J_updown}\\
J_z &= J_{1,z}\otimes 1 + 1\otimes J_{2,z}, \label{J_z}
\end{align}
where $1$ and $2$ in the subscript are used to specify the subsystem, and
\begin{equation}
k_i = q^{J_{i,z}}. \label{k}
\end{equation}

The quadratic Casimir operator for $U_q(sl_2)$ is
\begin{equation}
C = X_- X_+ + \left[J_z+\frac{1}{2}\right]_q^2,
\label{casimir}
\end{equation}
where
\begin{equation}
[n]_q = \frac{q^n-q^{-n}}{q-q^{-1}}. 
\label{n_q}
\end{equation}
The quantum numbers $J$ and $m$ are defined from the eigenvalues of $C$ and $J_z$:
\begin{align}
C\ket{J,m} &= \left[J+\frac{1}{2}\right]_q^2\ket{J,m}, \label{casimir ket}\\
J_z\ket{J,m} &= m\ket{J,m}. \label{Jz ket}
\end{align}
The ascending and descending operators act on the state $\ket{J,m}$ as
\begin{equation}
X_{\pm}\ket{J,m}=\sqrt{[J\mp m]_q[J\pm m+1]_q}\ket{J,m\pm 1}. \label{J_ket}
\end{equation}

In the same way as $SU(2)$, we can derive the recurrence relation to construct the many-spin eigenstates of the quantum group by using Eqs. (\ref{J_updown}) and (\ref{J_ket}). If we add a spin to the right side of a $N$-spin system we obtain
\begin{equation}
\begin{pmatrix}
\ket{J+\frac{1}{2},m}_{N+1} \\
\ket{J-\frac{1}{2},m}_{N+1} \\
\end{pmatrix}
=
\begin{pmatrix}
\sqrt{q^{-(J-m+\frac{1}{2})}\frac{[J+m+\frac{1}{2}]_q}{[2J+1]_q}} & \sqrt{q^{(J+m+\frac{1}{2})}\frac{[J-m+\frac{1}{2}]_q}{[2J+1]_q}} \\
-\sqrt{q^{(J+m+\frac{1}{2})}\frac{[J-m+\frac{1}{2}]_q}{[2J+1]_q}} & \sqrt{q^{-(J-m+\frac{1}{2})}\frac{[J+m+\frac{1}{2}]_q}{[2J+1]_q}} \\
\end{pmatrix}
\begin{pmatrix}
\ket{J,m-\frac{1}{2}}_N\otimes\ket{\uparrow} \\
\ket{J,m+\frac{1}{2}}_N\otimes\ket{\downarrow} \\
\end{pmatrix}
,
\label{recurrence_1 ago}
\end{equation}
where $\ket{J,m}_N$ represents a state of $N$-spin system. If we add a spin to the left, we obtain
\begin{equation}
\begin{pmatrix}
\ket{J+\frac{1}{2},m}_{N+1} \\
\ket{J-\frac{1}{2},m}_{N+1} \\
\end{pmatrix}
=
\begin{pmatrix}
\sqrt{q^{(J-m+\frac{1}{2})}\frac{[J+m+\frac{1}{2}]_q}{[2J+1]_q}} & \sqrt{q^{-(J+m+\frac{1}{2})}\frac{[J-m+\frac{1}{2}]_q}{[2J+1]_q}} \\
-\sqrt{q^{-(J+m+\frac{1}{2})}\frac{[J-m+\frac{1}{2}]_q}{[2J+1]_q}} & \sqrt{q^{(J-m+\frac{1}{2})}\frac{[J+m+\frac{1}{2}]_q}{[2J+1]_q}} \\
\end{pmatrix}
\begin{pmatrix}
\ket{\uparrow}\otimes \ket{J,m-\frac{1}{2}}_N\\
\ket{\downarrow}\otimes\ket{J,m+\frac{1}{2}}_N \\
\end{pmatrix}
.
\label{recurrence_2 ago}
\end{equation}
We can construct a state $\ket{J,m}_N$ by using Eqs. (\ref{recurrence_1 ago}) and (\ref{recurrence_2 ago}) repeatedly. Note that one can construct many different states corresponding to a certain set of the quantum numbers $J,m$ for a $N$-spin system, depending on the history of the composition of the state. The projection operator for $J,m$ is obtained by adding up all the projection operators for such states:
\begin{equation}
P_{N,J,m} = \sum_i\ket{J,m,i}_N{}_N\bra{J,m,i},
\label{projection sum}
\end{equation}
where $i$ is an index to specify the history of the composition of the state.

\subsection{Clebsch-Gordan Coefficients for the Quantum Group in the Limit $q\rightarrow0$}
Taking the limit $q\rightarrow0$, Eqs. (\ref{recurrence_1 ago}) and (\ref{recurrence_2 ago}) become very simple. For $-J+\frac{1}{2} \le m \le J-\frac{1}{2}$ we have
\begin{equation}
\begin{pmatrix}
\ket{J+\frac{1}{2},m}_{N+1} \\
\ket{J-\frac{1}{2},m}_{N+1} \\
\end{pmatrix}
=
\begin{pmatrix}
1 & 0 \\
0 & 1 \\
\end{pmatrix}
\begin{pmatrix}
\ket{J,m-\frac{1}{2}}_N\otimes\ket{\uparrow} \\
\ket{J,m+\frac{1}{2}}_N\otimes\ket{\downarrow} \\
\end{pmatrix}
,
\label{recurrence_1}
\end{equation}
\begin{equation}
\begin{pmatrix}
\ket{J+\frac{1}{2},m}_{N+1} \\
\ket{J-\frac{1}{2},m}_{N+1} \\
\end{pmatrix}
=
\begin{pmatrix}
0 & 1 \\
-1 & 0 \\
\end{pmatrix}
\begin{pmatrix}
\ket{\uparrow}\otimes \ket{J,m-\frac{1}{2}}_N\\
\ket{\downarrow}\otimes\ket{J,m+\frac{1}{2}}_N \\
\end{pmatrix}
.
\label{recurrence_2}
\end{equation}
For $m=\pm \left(J+\frac{1}{2}\right)$, we have
\begin{align}
\Ket{J+\frac{1}{2}, J+\frac{1}{2}}_{N+1} &=  \ket{J,J}_N\otimes\ket{\uparrow}, \label{edge 1} \\
\Ket{J+\frac{1}{2}, -J-\frac{1}{2}}_{N+1} &=  \ket{J,-J}_N\otimes\ket{\downarrow} \label{edge 2}, \\
\nonumber
\end{align}
\begin{align}
\Ket{J+\frac{1}{2}, J+\frac{1}{2}}_{N+1} &=  \ket{\uparrow}\otimes\ket{J,J}_N, \label{edge 3}\\
\Ket{J+\frac{1}{2}, -J-\frac{1}{2}}_{N+1} &=  \ket{\downarrow}\otimes\ket{J,-J}_N. \label{edge 4}\\
\nonumber
\end{align}

\section{Matrix Elements of the Heat Bath Superoperators}
\label{D_element}
In this appendix, we show how to derive analytic expressions for matrix elements of the heat bath superoperators. The recurrence relations for the projection operator derived from Eq. (\ref{recurrence_1 ago}) are
\begin{align}
P_{N+1,J+\frac{1}{2},m}=c_{J+\frac{1}{2},m,q}^2P_{N,J,m-\frac{1}{2}}\otimes\sigma_u
+s_{J+\frac{1}{2},m,q}^2P_{N,J,m+\frac{1}{2}}\otimes\sigma_d\nonumber\\
+c_{J+\frac{1}{2},m,q}s_{J+\frac{1}{2},m,q}(R_{N,m-\frac{1}{2},m+\frac{1}{2}}\otimes\sigma
_{+}+h.c.),
\label{recurrence_P_J+}
\end{align}
\begin{align}
P_{N+1,J-\frac{1}{2},m}=s_{J+\frac{1}{2},m,q}^2P_{N,J,m-\frac{1}{2}}\otimes\sigma_u
+c_{J+\frac{1}{2},m,q}^2P_{N,J,m+\frac{1}{2}}\otimes\sigma_d\nonumber\\
-c_{J+\frac{1}{2},m,q}s_{J+\frac{1}{2},m,q}(R_{N,m-\frac{1}{2},m+\frac{1}{2}}\otimes\sigma
_{+}+h.c.),
\label{recurrence_P_J-}
\end{align}
where
\begin{align}
c_{J,m,q}&=\sqrt{q^{-(J-m)}\frac{[J+m]_q}{[2J]_q}}, \label{c_jmq}\\
s_{J,m,q}&=\sqrt{q^{J+m}\frac{[J-m]_q}{[2J]_q}}, \label{s_jmq}\\ 
R_{N,J,m,m'}&=\sum_i\ket{J,m,i}\bra{J,m',i}, \label{R_Njm}\\   
\sigma_{+}&=\ket{\uparrow}\bra{\downarrow}. \label{sigma_plus}    
\end{align}
We define the matrix elements of the heat bath super operator $\mathcal{D}_l$ ($l=L, R$) as
\begin{equation}
\mathcal{D}_{l,N,J',m',J,m} = {\rm Tr}(P_{N,J',m'}\mathcal{D}_lP_{N,J,m}). 
\label{Dl_element}
\end{equation}
Since $\mathcal{D}_l$ acts only on the edge spin, the matrix element Eq. (\ref{Dl_element}) is zero if $|J'-J|>1$ or $|m'-m|>1$. For $|J'-J|\le1$ and $|m'-m|\le 1$, the matrix elements for $\mathcal{D}_R$ are the following.
\begin{align}
\mathcal{D}_{R,N+1,J,m,J,m}
&=\{(1+t_R)c^4_{J,m,q}+(1-t_R)s^4_{J,m,q}\}W_{N,J-\frac{1}{2}} \nonumber \\
&+\{(1+t_R)s^4_{J+1,m,q}+(1-t_R)c^4_{J+1,m,q}\}W_{N,J+\frac{1}{2}} \nonumber \\
&+2c^2_{J,m,q}s^2_{J,m,q}W_{N,J-\frac{1}{2}}\nonumber \\
&+2c^2_{J+1,m,q}s^2_{J+1,m,q}W_{N,J+\frac{1}{2}}, \label{D_jmjm}
\end{align}
\begin{align}
\mathcal{D}_{R,N+1,J+1,m,J,m} 
=& 0,\label{D_j+1mjm}
\end{align}
\begin{align}
\mathcal{D}_{R,N+1,J-1,m,J,m}
=&0,\label{D_j-1mjm}
\end{align}
\begin{align}
\mathcal{D}_{R,N+1,J,m+1,J,m}
=&-(1-t_R)c^2_{J,m+1,q}s^2_{J,m,q}W_{N,J-\frac{1}{2}} \nonumber\\
&-(1-t_R)s^2_{J+1,m+1,q}c^2_{J+1,m,q}W_{N,J+\frac{1}{2}},\label{D_jm+1jm}
\end{align}
\begin{align}
\mathcal{D}_{R,N+1,J,m-1,J,m}
=&-(1+t_R)s^2_{J,m-1,q}c^2_{J,m,q}W_{N,J-\frac{1}{2}}\nonumber\\
&-(1+t_R)c^2_{J+1,m-1,q}s^2_{J+1,m,q}W_{N,J+\frac{1}{2}}, \label{D_jm-1jm}
\end{align}
\begin{align}
\mathcal{D}_{R,N+1,J+1,m+1,J,m}
=&-(1-t_R)c^2_{J+1,m+1,q}c^2_{J+1,m,q}W_{N,J+\frac{1}{2}}, \label{D_j+1m+1jm}
\end{align}
\begin{align}
\mathcal{D}_{R,N+1,J-1,m+1,J,m}
=&-(1-t_R)s^2_{J,m+1,q}s^2_{J,m,q}W_{N,J-\frac{1}{2}}, \label{D_j-1m+1jm}
\end{align}
\begin{align}
\mathcal{D}_{R,N+1,J+1,m-1,J,m}
=&-(1+t_R)s^2_{J+1,m-1,q}s^2_{J+1,m,q}W_{N,J+\frac{1}{2}}, \label{D_j+1m-1jm}
\end{align}
\begin{align}
\mathcal{D}_{R,N+1,J-1,m-1,J,m}
=&-(1+t_R)c^2_{J,m-1,q}c^2_{J,m,q}W_{N,J-\frac{1}{2}}, \label{D_j-1m-1jm}
\end{align}
where
\begin{equation}
W_{N,J}=\frac{4J+2}{N+2J+2}{}_{N}C_{\frac{N}{2}-J}
\label{dimension}
\end{equation}
is the dimension of the projection operator $P_{N,J,m}$. Matrix elements of $\mathcal{D}_L$ is obtained by setting $q\rightarrow q^{-1}$ and $t_{R}\rightarrow t_{L}$ in the above expressions for $\mathcal{D}_R$. Finally, matrix representation of $\mathcal{L}_2 = \mathcal{D}_L + \mathcal{D}_R$ is obtained by adding the matrices for $\mathcal{D}_L$ and $\mathcal{D}_R$.

\section{Recurrence Relations for Some Physical Quantities of the Projection Operators}
\label{recursion}

\subsection{Local Temperature}
We define the local temperature parameter for the projection operator $P_{N,J,m}$ by
\begin{equation}
t_{N,n,J,m} = -\frac{1}{W_{N,J}}{\rm Tr}\left(\sigma_n^z P_{N,J,m}\right). 
\label{t_Jm}
\end{equation}
The expectation value for the steady state is obtained as the weighted average
\begin{equation}
t_{n}=\sum_{J,m}C_{J,m}t_{N,n,J,m},    
\label{t_n sum}
\end{equation}
where $C_{J,m}$'s are weight coefficients for the density matrix.

The local temperature parameter for the projection operator satisfies the following recurrence relation
\begin{eqnarray}
t_{N,n,J,m}
&=&
\frac{J(N+2J+2)}{(2J+1)N}\left(
c_{J,m,q}^2t_{N-1,n,J-\half,m-\half} +
s_{J,m,q}^2t_{N-1,n,J-\half,m+\half}\right)\nonumber\\
&&
+ \frac{(J+1)(N-2J)}{(2J+1)N}
\left(
s_{J+1,m,q}^2 t_{N-1,n,J+\half,m-\half} +
c_{J+1,m,q}^2t_{N-1,n,J+\half,m+\half}\right)
\label{t_recurrence}
\end{eqnarray}
for $1\le n \le N-1$.

For $n=N$ we can use another relation
\begin{eqnarray}
t_{N,N,J,m}
&=&
\frac{J(N+2J+2)}{(2J+1)N}\left(
c_{J,m,q}^2 -s_{J,m,q}^2\right)+\frac{(J+1)(N-2J)}{(2J+1)N}
\left(
s_{J+1,m,q}^2 - c_{J+1,m,q}^2\right).
\label{t_recurrence_edge}
\end{eqnarray}

\subsection{Correlation Function}
We define the $z$-$z$ spin correlation function for the projection operator $P_{N,J,m}$ by 
\begin{equation}
S_{N,J,m}=\frac{1}{W_{N,J}}{\rm Tr}(\sigma_{l}^z\sigma_{r}^zP_{N,J,m}).
\label{S_NJm}
\end{equation}
The correlation function for the steady state is obtained as the weighted average of $S_{N,J,m}$'s in the same way as in the case of the local temperature. 

The correlation function Eq. (\ref{S_NJm}) satisfies the following recurrence relation
\begin{eqnarray}
S_{N,J,m} 
&=&
\frac{J(N+2J+2)}{(2J+1)N}
\left(c_{J,m,q}^{2}S_{N-1,J-\half,m-\half}+
s_{J,m,q}^{2}S_{N-1,J-\half,m+\half} \right) \nonumber \\
&&
+ \frac{(J+1)(N-2J)}{(2J+1)N}
\left(
s_{J+1,m,q}^{2}S_{N-1,J+\half,m-\half+}
c_{J+1,m,q}^{2}S_{N-1,J+\half,m+\half} \right)
\label{S_N+1Jm}
\end{eqnarray}
for $N> r$. 

For $N=r$ we can use another relation
\begin{eqnarray}
S_{r,J,m} 
&=&
\frac{J(r+2J+2)}{(2J+1)r}
\left(
c_{J,m,q}^{2}t_{r-1,l,J-\half,m-\half}-
s_{J,m,q}^{2}t_{r-1,l,J-\half,m+\half}\right) \nonumber \\
&&
+\frac{(J+1)(r-2J)}{(2J+1)r}
\left(
s_{J+1,m,q}^{2}t_{r-1,l,J+\half,m-\half}-
c_{J+1,m,q}^{2}t_{r-1,l,J+\half,m+\half}\right).
\label{S_rJm}
\end{eqnarray}

We define the $x$-$x$ spin correlation function for the projection operator $P_{N,J,m}$ by 
\begin{align}
U_{N,J,m}=\frac{1}{W_{N,J}}{\rm Tr}(\sigma_{l}^x\sigma_{r}^xP_{N,J,m}).
\label{U_NJm}
\end{align}
Then it satisfies the following recurrence relation
\begin{align}
U_{N,J,m} 
=&
\frac{J(N+2J+2)}{(2J+1)N}
\left(
c_{J,m,q}^{2}U_{N-1,J-\half,m-\half}+s_{J,m,q}^{2}U_{N-1,J-\half,m+\half}\right) \nonumber \\
&+\frac{(J+1)(N-2J)}{(2J+1)N}
\left(
s_{J+1,m,q}^2U_{N-1,J+\half,m-\half} +
c_{J+1,m,q}^2U_{N-1,J+\half,m+\half}\right)
\label{U_N+1Jm}
\end{align}
for $N>r$.

For $N=r$ we can use another relation
\begin{align}
U_{r,J,m} 
=&
\frac{J(r+2J+2)}{(2J+1)r} c_{J,m,q} s_{J,m,q}
\left(
V_{r-1,J-\half,m-\half,m+\half}+V_{r-1,J-\half,m+\half,m-\half}\right) \nonumber \\
&-\frac{(J+1)(r-2J)}{(2J+1)r}c_{J+1,m,q} s_{J+1,m,q}
\left(
V_{r-1,J+\half,m-\half,m+\half} + V_{r-1,J+\half,m+\half,m-\half}\right),
\label{U_rJm}
\end{align}
where  
\begin{align}
V_{n,J,m,m^{'}}=\frac{1}{W_{N,J}}{\rm Tr}(\sigma_{l}^x R_{n,J,m,m^{'}}).
\label{V_nJm}
\end{align}
It satisfies the following recurrence relation
\begin{align}
V_{n,J,m,m^{'}}
=&\frac{J(n+2J+2)}{(2J+1)n}
\left(c_{J,m,q}c_{J,m^{'},q}V_{n-1,J-\half,m-\half,m^{'}-\half}+s_{J,m,q}s_{J,m^{'},q}V_{n-1,J-\half,m+\half,m^{'}+\half} \right) \nonumber \\
&+\frac{(J+1)(n-2J)}{(2J+1)n}
\left(s_{J+1,m,q} s_{J+1,m^{'},q} V_{n-1,J+\half,m-\half,m^{'}-\half}+c_{J+1,m,q}c_{J+1,m^{'},q}V_{n-1,J+\half,m+\half,m^{'}+\half}\right)
\label{V_n+1Jm}
\end{align}
for $l < n \leq r-1$.

For $n=l$ we can use another relation
\begin{align}
V_{l,J,m,m^{'}}
=&
\frac{J(l+2J+2)}{(2J+1)l}c_{J,m,q}s_{J,m^{'},q}
\left(
\delta_{m-\half,m^{'}+\half}+\delta_{m^{'}+\half,m-\half}
\right) \nonumber \\
&-\frac{(J+1)(l-2J)}{(2J+1)l}s_{J+1,m,q}c_{J+1,m^{'},q}
\left(
\delta_{m-\half,m^{'}+\half}+\delta_{m^{'}+\half,m-\half}
\right).
\label{V_lJm}
\end{align}

\section{Derivation of the Equation(\ref{j}) in Detail}
\label{Energy current}
The system Hamiltonian in this case is
\begin{align}
H_S&=H_0+\lambda H_I\thickapprox H_0 \label{H_s_appro}\\
&=h\sum_{l=1}^N\sigma_l^z.\label{H_s_sigma}
\end{align}
Then, substituting Eq. (\ref{H_s_sigma}) into Eq. (\ref{j before}), we obtain
\begin{align}
j&=-h\sum_{l=1}^N\sum_j\operatorname{Tr}(\lbrack\sigma_j,S_j\rho-\rho S_j^\dagger\rbrack\sigma_l^z) \label{j_appen ago}\\
&=-h\operatorname{Tr}((D_{L}\rho)\sigma_1^z). \label{j_appen}
\end{align}
Since the heat bath superoperator acts only on the edge spin, we represent the density matrix as
\begin{equation}
\rho=\sigma_1^0\otimes \rho^0+\sigma_1^x\otimes \rho^x+\sigma_1^y\otimes \rho^y+\sigma_1^z\otimes \rho^z.
\label{rho_edge}
\end{equation}
Applying $\mathcal{D}_L$ on Eq. (\ref{rho_edge}), we obtain 
\begin{equation}
\mathcal{D}_L\rho=at_L\sigma_1^z\otimes \rho^0+b\sigma_1^x\otimes\rho^x+b\sigma_1^y\otimes\rho^y+a\sigma_1^z\otimes\rho^z.
\label{DLrho}
\end{equation}
Then, substituting Eq. (\ref{DLrho}) into Eq. (\ref{j_appen}), we obtain
\begin{align}
j&=-h\lparen 
at_L\operatorname{Tr}(\sigma_1^0)\operatorname{Tr}(\rho^0)+a\operatorname{Tr}(\sigma_1^0)\operatorname{Tr}(\rho^z)\rparen \label{DLrhosigma ago}\\
&=ah\lparen t_1-t_L\rparen. \label{DLrhosigma}
\end{align}
Thus we obtain Eq. (\ref{j}).


\begin{thebibliography}{9}
\bibitem{Lepri}S. Lepri, R. Livi, and A. Politi, Phys. Rep. 377, 1 (2003).
\bibitem{Prosen}T. Prosen and M. \v{Z}nidari\v{c}, J. Stat. Mech. P02035 (2009).
\bibitem{Michel}M. Michel, O. Hess, H. Wichterich, and J. Gemmer, Phys. Rev. B 77, 104303 (2008).
\bibitem{Wu}J. Wu and M. Berciu, Phys. Rev. B 83, 214416 (2011).
\bibitem{Mendoza}J. J. Mendoza-Arenas, S. Al-Assam, S. R. Clark, and D. Jaksch, J. Stat. Mech. P07007 (2013).
\bibitem{Yuge} T. Yuge and A. Sugita, J. Phys. Soc. Jpn. 84, 014001 (2014).
\bibitem{Sugita}A. Sugita, Proc. Recent Advances in the Physics of Low-Dimensional Nanoscale Systems,
arXiv:1203.3817.
\bibitem{Ishida}T. Ishida and A. Sugita, J. Phys. Soc. Jpn. 85, 074006 (2016)
\bibitem{tdmrg}A. J. Daley, C. Kollath, U. Schollw\"{o}ck and G. Vidal, J. Stat. Mech. (2004) P04005
\bibitem{Saito}K. Saito, S. Takesue, and S. Miyashita, Phys. Rev. E 61, 2397 (2000).

\end{thebibliography}
\end{document}